\documentclass[a4paper,fleqn,usenatbib,useAMS]{mnras}

\usepackage{graphicx}	
\usepackage{amsmath}	
\usepackage{amssymb}	
\usepackage{multicol}        
\usepackage{bm}		
\usepackage{pdflscape}	
\usepackage[T1]{fontenc}
\usepackage{ae,aecompl}

\usepackage{txfonts}

\title[Million-Body Comparisons between Monte Carlo and Direct $N$-body]{Million-Body Star Cluster Simulations: Comparisons between Monte Carlo and Direct $N$-body}

\author[C. L. Rodriguez et al.]{Carl L. Rodriguez$^{1,2}$\thanks{Contact e-mail: \href{mailto:cr@u.northwestern.edu}{cr@u.northwestern.edu}}, Meagan Morscher$^{1,2}$, Long Wang$^{3,4}$, Sourav Chatterjee$^{1,2}$,  \newauthor Frederic A. Rasio$^{1,2}$, Rainer Spurzem$^{3,5,6,7}$
\\
$^{1}$Center for Interdisciplinary Exploration and Research in Astrophysics (CIERA), Northwestern University, Evanston, IL, USA\\
$^{2}$Department of Physics and Astronomy, Northwestern University, Evanston, IL, USA\\
$^{3}$Kavli Institute for Astronomy and Astrophysics, Peking University, Yiheyuan Lu 5, Haidian Qu, 100871, Beijing, China\\
$^{4}$Department of Astronomy, School of Physics, Peking University, Yiheyuan Lu 5, Haidian Qu, 100871, Beijing, China\\
$^{5}$National Astronomical Observatories and Key Laboratory of Computational Astrophysics, Chinese Academy of Sciences, \\~~~~~~20A Datun Rd., Chaoyang District, 100012, Beijing, China\\
$^{6}$Key Laboratory of Frontiers in Theoretical Physics, Institute of Theoretical Physics, Chinese Academy of Sciences, Beijing, 100190, China\\
$^{7}$Astronomisches Rechen-Institut, Zentrum f{\"u}r Astronomie, University of Heidelberg, M{\"o}nchhofstrasse 12-14,
69120, Heidelberg, Germany}

\date{Last updated \today}

\pubyear{2016}

\begin{document}
\label{firstpage}
\pagerange{\pageref{firstpage}--\pageref{lastpage}}
\maketitle

\begin{abstract}
We present the first detailed comparison between million-body globular cluster simulations computed with a H\'enon-type 
Monte Carlo code, CMC, and a direct $N$-body code, NBODY6++GPU. Both simulations 
start from an identical cluster model with $10^6$ particles, 
and include all of the relevant physics needed to treat the system in a highly realistic way. 
With the two codes ``frozen" (no fine-tuning of 
any free parameters or internal algorithms of the codes) we find excellent agreement in the overall evolution of the two models. Furthermore,
we find that in both models, large numbers of stellar-mass black holes ($> 1000$) are retained for 12 Gyr. Thus, the very accurate direct $N$-body approach
confirms recent predictions that black holes can be retained in present-day, old globular clusters. We find only minor disagreements between the two models and attribute these to the small-$N$ dynamics driving the evolution of the cluster core for which the Monte Carlo assumptions are less ideal.  Based on the overwhelming general agreement between the two models computed using these vastly different techniques, 
we conclude that our Monte Carlo approach, which is more approximate, but dramatically faster compared to the direct $N$-body, is capable of producing a very accurate description 
of the long-term evolution of massive globular clusters even when the clusters contain large populations of stellar-mass black holes. 
\end{abstract}


\section{Introduction} \label{Intro}

In the last few years, our understanding of the dynamical evolution of globular clusters (GC), especially, as a result of the detailed dynamical evolution and fate of the large numbers of black holes (BHs) that are bound to form in these large-$N$ clusters, 
has shifted significantly. While it was once thought
that present-day old GCs should have at most a couple of BHs remaining, more recent studies have shown
that if BH-formation kicks are not sufficiently high to eject all BHs from the GCs, a significant fraction of the formed BHs are retained up to the typical old ages ($\sim 12$ Gyr) of 
the GCs
\citep{Breen2013, Heggie2014, Mackey2008, Sippel2013, Morscher2015}. This new 
perspective has been driven by major advances in parallel computing, which enables the modeling of GC systems with realistic number of stars and all the relevant physics.

Since the early theoretical work predicting the rapid ejection of BHs from clusters, several groups have
performed ever more realistic evolutionary simulations. While several of the first attempts confirmed the
theoretical prediction for complete evaporation, the most recent models have predicted that at least some BHs,
and likely many, may remain in old GCs today. These results are coming at an exciting time when
stellar BHs are being discovered at a rapid pace in GCs in the Milky Way and in other Galaxies through X-ray 
and radio surveys \citep{MaccaroneNature2007, Barnard2011, Maccarone2011, Shih2010, Strader2012, Chomiuk2013}.

We are also beginning to understand why these new results are at odds with the early theoretical prediction
of rapid BH evaporation. The original argument assumed that the BHs would become Spitzer unstable due to their rapid mass-segregation,
 dynamically decoupling from the cluster.  Once decoupled, the BHs would evolve as an isolated cluster, with an evaporation timescale of $\sim1$ Gyr.
While there were a few theoretical hints from simulations that at least some BHs could be retained for 
12 Gyr, there was no good explanation for how this could be possible, and where the old prediction 
breaks down. Using two-component direct $N$-body models, \cite{Breen2013} showed that after the BHs 
began to segregate, their rate of energy generation was controlled by the rate of energy flow through the 
cluster as a whole,
and so it was set by the relaxation timescale of the entire cluster, not simply by that of the small subpopulation of BHs.
Recently, \citet{Morscher2015} (hereafter MOR15) presented a grid of 42 detailed Monte Carlo models 
for realistic star clusters containing substantial initial populations of stellar BHs. This study was the first to present a large number of realistic simulations with a range of initial cluster masses (from $\sim 10^5-10^6\, M_{\odot}$) as well as
variation in other important parameters, such as the virial radius, that also includes all of the relevant physics required to describe these systems accurately. 
Starting from standard assumptions for the initial conditions of MW-like clusters and BH formation processes, 
MOR15 found that nearly all of the models did indeed retain significant fractions of their BHs up to the end of 
the simulation at 12 Gyr. The retained BHs heated the full cluster, leading to large final core sizes; however, the most compact initial cluster models were able to eject the majority of their BHs by 12 Gyr, causing their cores to contract (the so-called second core collapse) to sizes similar to those observed in MWGCs. Work is in progress to study the effect of varying the initial mass function and the BH birth kick distribution, on the initial BH populations in clusters, and ultimately their impact on cluster dynamics and long-term BH retention.

Perhaps most interestingly, MOR15 has revealed a new theoretical understanding of the complex dynamical interplay 
between BHs and clusters, and how it may be possible for clusters to retain large numbers of BHs for many Gyr. 
These MC models showed that, in realistic systensm, the Spitzer instability does \emph{not} involve all of the BHs in the cluster.
Rather, the BHs power core oscillations during which only a small subset of the BHs segregate 
from the cluster to form a deep cusp, which then promptly re-expands upon formation of three-body binaries. 
 Thus, as a result of the energy generated by the dynamics of the BHs, the BH interaction rate, and therefore also the 
 evaporation rate, is kept much lower than previously thought, making it possible for star clusters to retain significant 
 numbers of BHs for 12 Gyr.  
It seems that we should not expect the BHs to succomb to the Spitzer Instability;
most of the BHs, in fact, \emph{always} remain spread throughout the cluster, far from the central cusp.
This new understanding of BH dynamics in star clusters has provided a theoretical basis for explaining
the recent discoveries of several BH candidates in old GCs.

As always, we would like to be able to compare our results to simulations done with other codes, especially those that use different
modeling techniques. The direct $N$-body technique is the most accurate and assumption-free method for modeling
the dynamics of star clusters. It resolves physics on the dynamical timescale, which means it can in principle be used to 
model clusters of any size, because it is valid even for clusters with relaxation timescales not much longer than their 
dynamical timescales. This is in contrast to the MC technique, whose assumptions rely on the relaxation timescale being 
long compared to the dynamical timescale. For large enough $N$, however, the MC technique used in our study is an
extremely powerful tool that is capable of computing \emph{many} large-$N$ dynamical cluster models in a short amount 
of time, which means it can be used to explore the parameter space of initial conditions and test the robustness of our results. 
This feature sets it apart from direct $N$-body simulations, which are significantly
more computationally expensive, and therefore are usually restricted to $N \lesssim 10^5$. 
The speed of the MC technique
comes at a price, however, in that two-body relaxation, as well as other dynamical processes, are treated in an approximate 
way via a single interaction per pair of stars per time step, where the timestep is chosen to be a small fraction of the relaxation 
timescale. Physics occuring on much shorter timescales, such as the dynamical timescale, cannot be resolved accurately 
with the MC method, but is handled naturally with the direct $N$-body technique.

Our MC code has been compared extensively
to direct $N$-body simulations whenever possible, and has shown excellent agreement (\citealt{Joshi2000, Joshi2001, 
Fregeau2003, Fregeau2007, Chatterjee2010, Umbreit2012}). However, the types
of clusters we are now capable of simulating are quite different than those we have simulated (and thus tested) in the past. In particular,
the inclusion of large populations of BHs with a broad spectrum of masses has a significant effect on cluster evolution,
producing clusters at 12 Gyr with very different properties than models without such objects (cf. \citealt{Chatterjee2010}, 
in which the BH mass spectrum was truncated at just above $2\, M_\odot$). 
For example, the deep BH-driven core oscillations that occur ubiquitously in our new CMC models are a new phenomenon
that we have not seen before in previous models. During these collapses, the physics may be governed by a very small number 
of objects ($\sim10$) that are partially decoupled from the rest of the cluster. If the relaxation timescale of the subset of stars is 
much smaller than the relaxation timescale of the cluster as a whole, then its evolution is perhaps not being computed accurately 
by our approximate MC scheme. Furthermore, it is possible that our crude prescription for forming three-body binaries is not capable
of properly capturing the physics of this dynamically important process.

These limitations of our MC calculations could potentially have an impact on our 
results regarding the dynamical evolution of BHs in dense clusters, including long-term BH retention and the resulting heating 
that produces puffy cluster cores. For these reasons, it would be desirable to re-test the results produced by the latest version 
of our MC code with the best currently-available direct $N$-body simulations.
For our specific interests, there have previously been no suitable large-$N$ direct $N$-body 
simulations available to which we could compare our large-$N$ MC simulations. 
Recently, however, \citealt{Wang2015} developed a new direct $N$-body code that employs hybrid parallelization 
methods to speed up the NBODY6++ code (\citealt{Spurzem1999, Spurzem2008}), which is considered to be the gold 
standard in direct $N$-body modeling. This new code, NBODY6++GPU, is capable of modeling clusters with $10^6$ 
stars within a year (\citealt{Wang2015,Wang2016}). This provides an excellent opportunity
for us to test our predictions regarding the evolution of clusters with BHs through a direct comparison.

Here we present a comparison between models generated with our MC code
(the \texttt{MC} model) and with NBODY6++GPU (the \texttt{NB} model). 
We present this as an honest comparison using what are believed to be the best versions of the two codes, and we do
not do any fine-tuning of free parameters. 
This rest of this paper is organized as follows. In Section~\ref{Method} we describe the two codes in more detail 
and give the initial conditions for the model that is the focus of this comparison.
In Section~\ref{MC_NB_Comparison} we present and compare the results of the two simulations, including the global
structural evolution as well as the evolution of the BH populations. We discuss some of the differences between the models and uncertainties and compare to other studies in Section~\ref{Discussion}. Finally
in Section~\ref{SummaryConclusions} we summarize our results and state our conclusions.


\section{Numerical Setup} \label{Method}

We compare the results of a small set of MC models to a single direct $N$-body simulation, 
all starting with identical initial conditions. The comparison model is a million-particle 
direct $N$-body simulation that has been 
computed with NBODY6++GPU (\citealt{Wang2015,Wang2016}), a newly developed optimized version of NBODY6++ with 
improved hybrid parallelization methods (MPI, GPU, OpenMP and AVX/SSE).

This new code combines the MPI parallelized NBODY6++
\citep{Spurzem1999,Hemsendorf2003} with the GPU and AVX/SSE libraries of
\citealt{Nitadori2012} to significantly accelerate the direct $N$-body
integratcion.  
They have also made improvements to speed up the time-step scheduling and 
stellar evolution, which had become bottlenecks after the hybrid parallelization scheme was implemented.
With these modifications, for cluster models consisting of single objects only, distributed 
over 320 CPU cores (across 16 nodes), with 86016 GPU cores (on 32 GPUs), they achieve a speed-up factor of 400-2000, 
depending on the number of stars. 
NBODY6++GPU is the first direct $N$-body code capable of simulating the evolution of a million-body collisional 
star cluster over many Gyr. For a detailed description of this hybrid code see \cite{Wang2015}.

Wang et al. have shared the results for the million-body simulation presented in \cite{Wang2016}. 
Provided with the exact initial cluster model used in their $N$-body simulation, we have performed a simulation
with our own MC code, CMC, which has been 
described in great detail in several earlier papers \citep{Joshi2000, Joshi2001, Fregeau2003, Fregeau2007, Chatterjee2010, Umbreit2012}. Briefly, we use a variation of the so-called 
``orbit-averaged Monte Carlo method" developed by \cite{Henon1971a} for solving the Fokker-Planck 
equation. With our recently parallelized MC code \citep{Pattabiraman2013}, we can calculate million-body simulations
such as the one presented here in less than 2 days\footnote{This is true for models with relatively large initial half-mass radii,
as is the case here. However, MC simulations starting from compact models with half mass radii of about 1 pc can take more 
than a month.} when distributed over just 48 CPU cores.

While these two codes employ very different techniques for modeling stellar dynamics, they include nearly all of the 
same physical processes, including two-body relaxation (treated in an approximate way by the MC code), direct physical 
collisions, higher-order strong binary encounters (integrated directly in both
codes), and single and interacting binary stellar 
evolution (using SSE and BSE, \citealt{Hurley2000,Hurley2002}). These stellar evolution prescriptions have been modified
from the original publicly available versions.  In the original SSE/BSE codes, only very low-mass BHs were formed. This was pointed out by \cite{Belczynski2002}, who also provided a new prescription for forming more realistic BH masses including fallback based on \cite{Fryer2001}. In both CMC
and NBODY6++GPU, the remnant masses are selected following the metallicity-dependent prescription of \cite{Belczynski2002} which forms BHs in the range of about $3-30\, M_\odot$ for typical GC metallicities.
Also, the magnitudes of the natal kicks received by stellar remnants that form via a supernova should
depend on the amount of fallback material, and both codes have also updated their birth kick prescriptions to depend
on fallback according to \cite{Belczynski2002}. The kicks are drawn from a Maxwellian velocity distribution with 
$\sigma=265$~km s$^{-1}$, then reduced proportionally to the mass of the material that falls back onto the newly-formed compact object.

\begin{figure*}
\centering
\includegraphics[trim=0cm 1cm 1cm 0cm,scale=0.73]{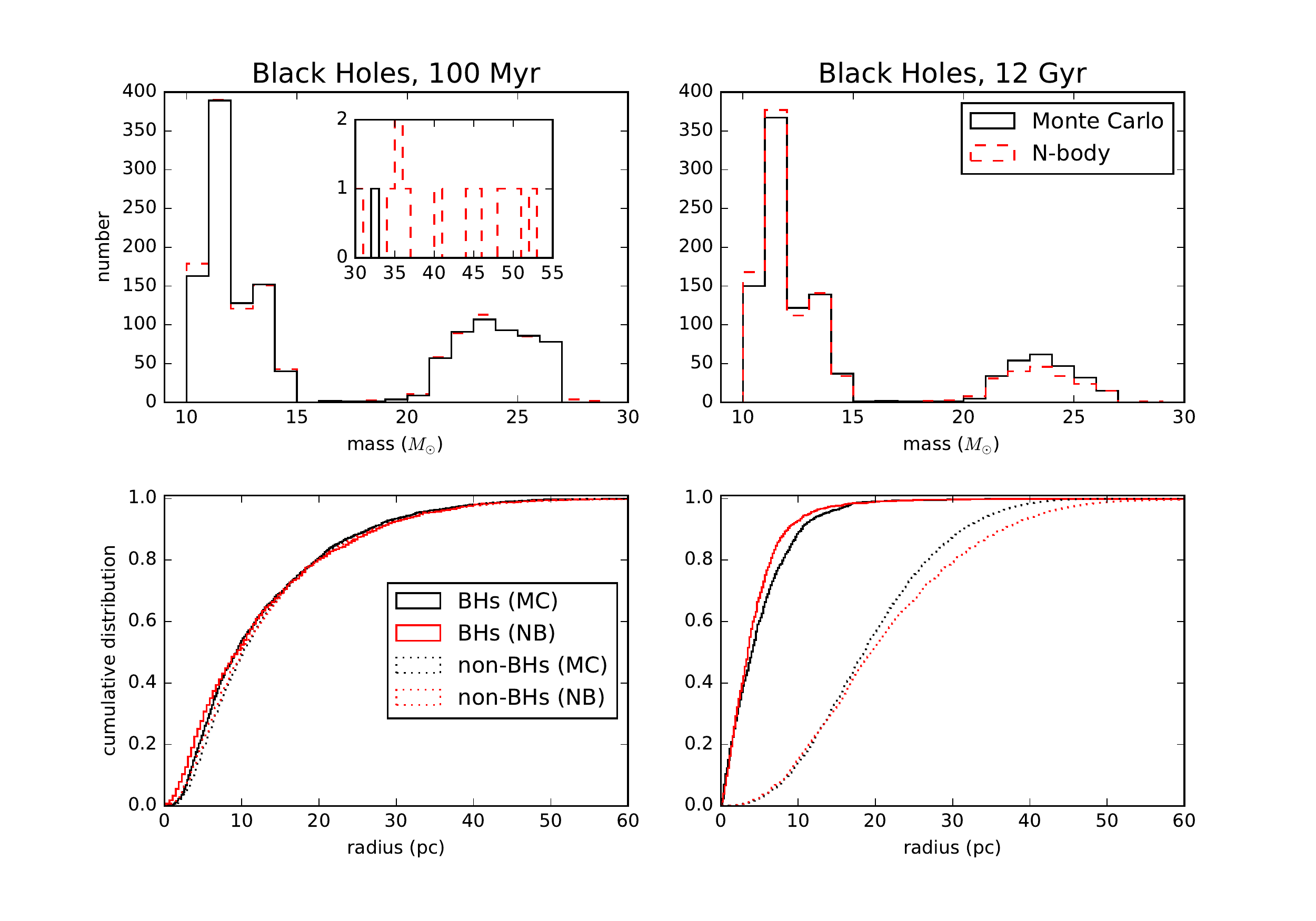}
	\caption{ The BH mass (top) and radial distributions (bottom) for BHs in both cluster models at 100 Myr (left) and 12 Gyr (right).
	On the top, the BH masses for model \texttt{MC} are shown by the solid black line, and those for model 
	\texttt{NB} are indicated by the dashed red line. Overall the number of retained BHs and the 
	mass distributions agree very well. There are a small number of BHs with masses between $30 M_{\odot}$ and $55 M_{\odot}$, highlighted in the insert, which are only formed in the \texttt{NB} model. The bottom panels show the radial distributions for the BHs (solid curves) and the
	non-BHs (dotted curves) separately, with model \texttt{MC} in black and model \texttt{NB} in red. 
	Each star is counted individually, regardless of whether it is a part of a binary (or an even higher-order system, 
	which is possible in the direct $N$-body simulation). }
	 \label{fig:bhmass_hist_final}
\end{figure*}

\begin{figure}
\centering
\includegraphics[trim=1cm 0cm 0cm 0cm,scale=0.45]{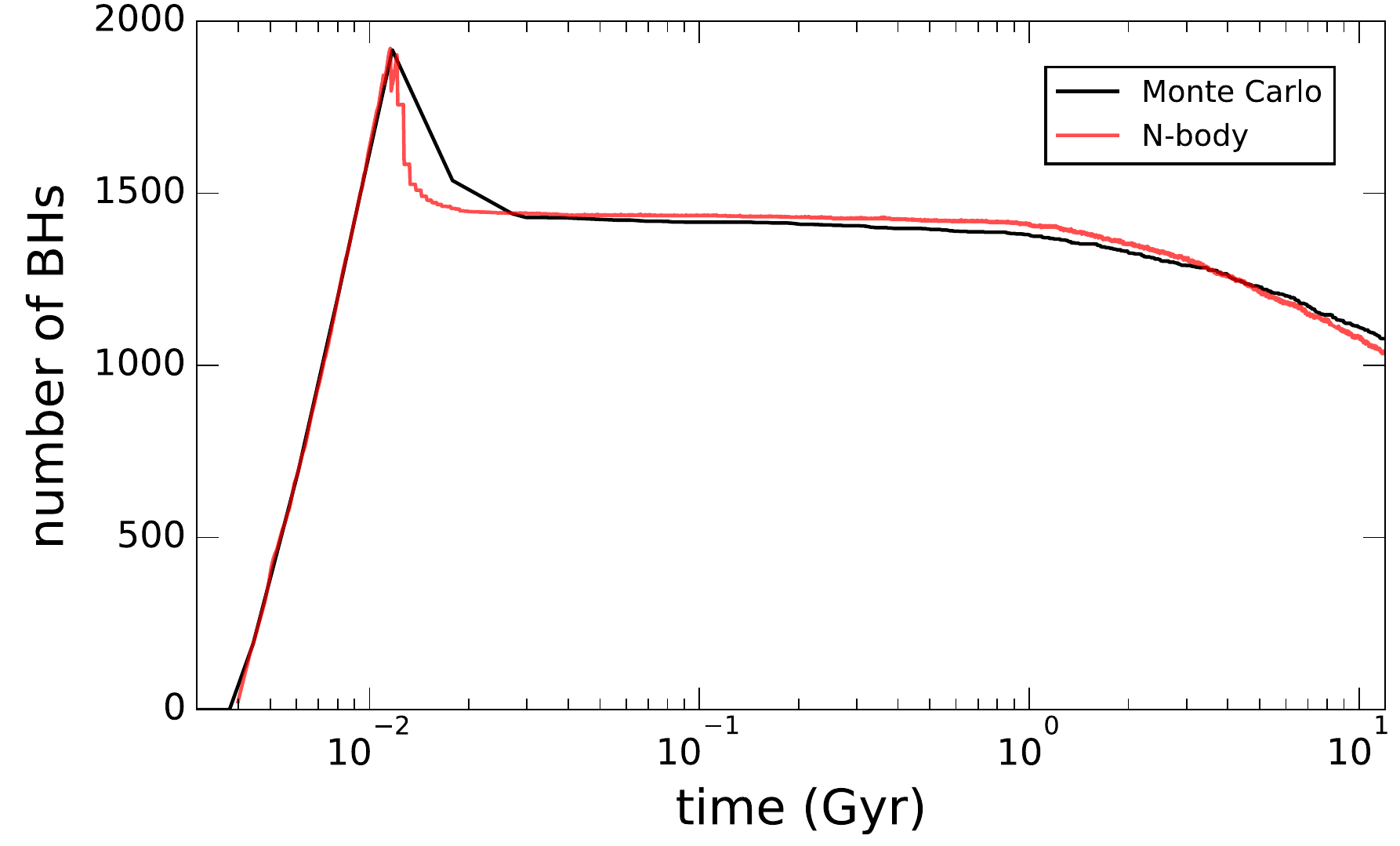}
 	 \caption{ Comparison of the total number of BHs retained in each cluster model 
	 		as a function of time. Model \texttt{MC} is shown in black and model \texttt{NB} in red.
			As already seen in Figure \ref{fig:bhformation}, both models forms and retains initially
			(at 20 Myr, after BH formation) a roughly identical number of BHs. Over time, BHs
			are slowly ejected in both models at a similar rate, with model \texttt{MC} ejecting 336 BHs and model \texttt{NB} ejecting 400 BHs by 12 Gyr.}
	\label{fig:Nbh_comp_LW}
\end{figure}

\begin{figure}
\centering
\includegraphics[trim=0cm 0cm 0cm 0cm,scale=0.81]{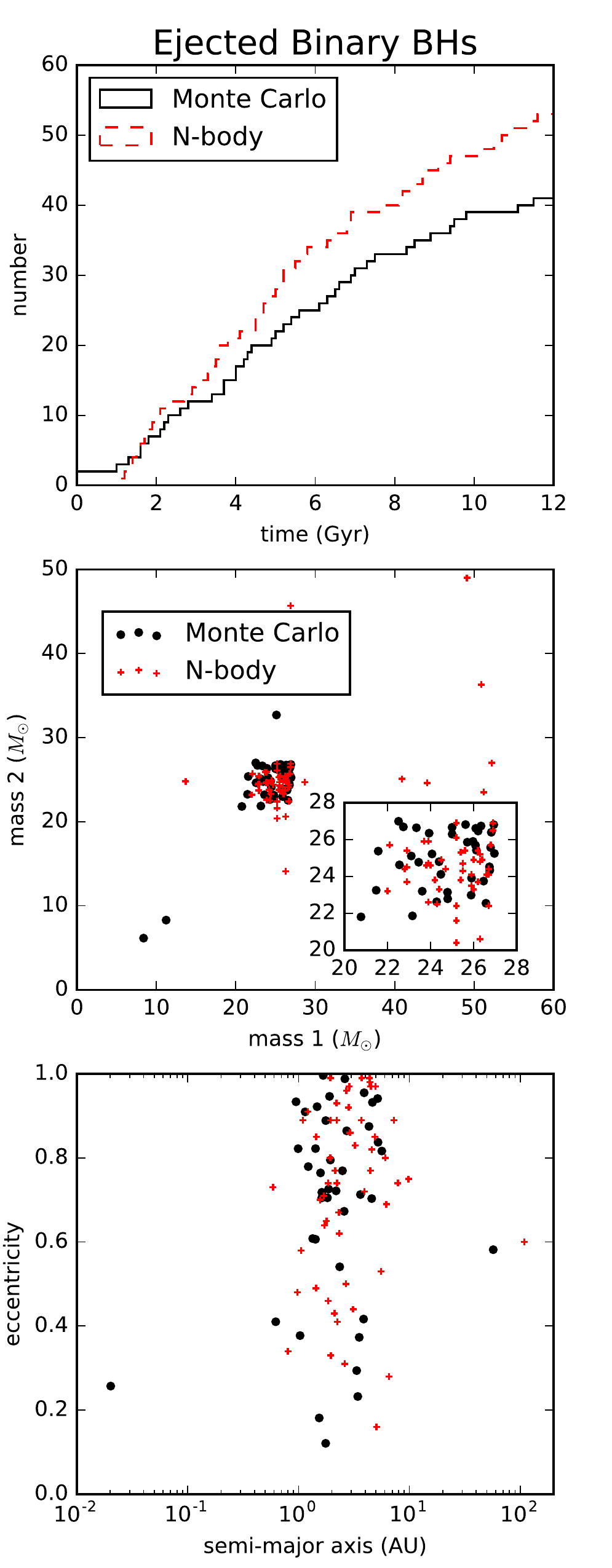}
 	 \caption{The properties of the binary BHs ejected from each simulation.  The top plot shows the number of ejected binaries over time for the \texttt{MC} model (solid black) and the \texttt{NB} model (dashed red).  The middle plot shows the component masses of the binaries (\texttt{MC} in black circles, \texttt{NB} in red crosses), while the bottom plot shows the eccentricity and semi-major axis distributions of the binaries (with same color scheme).  
	 }
	\label{fig:bhformation}
\end{figure}

While three-body
binary formation occurs naturally in direct $N$-body calculations, the MC technique relies on pair-wise
interactions between two neighboring particles, precluding the dynamical formation of a binary from encounters involving three particles. Therefore we opt to use a simple analytic prescription that relies on an estimate of the local probability of 
binary formation, as described in detail in MOR15. In the MC code, we use the exact physical assumptions and parameters as presented in MOR15 (e.g., those associated with three-body binary formation; see
Section 2.2, Equations 1 and 2 from MOR15).
One difference is that the MC code is not able to treat stable triple systems that form through four-body (binary-binary) encounters.
These systems are thus broken upon formation. However, since they form relatively rarely, we would not expect these to
have a significant impact on the evolution of the cluster. 

The initial conditions used in the model of \cite{Wang2016} were selected to be
similar to large GCs in the Milky Way.
They use a King model with concentration $W_o=6$, $N=10^6$, half-mass radius $R_h=7.5$ pc, binary fraction 
of 5\%, Galactocentric distance $R_G=7.1$ kpc, and metallicity $Z=0.00016$.
Stellar masses were chosen in the range $0.1 \,-\, 100\, M_\odot$ according to the initial mass 
function (IMF) of \cite{Kroupa2001}, which is a broken power-law of the form 
$dN/dm \propto m^{-\alpha}$, with $\alpha=1.3$ for $0.08\, \le\,  m / M_\odot \,\textless \,  0.5$, 
and $\alpha=2.3$ for $m / M_\odot \,\ge \, 0.5$. Binaries are created by choosing an existing single star 
randomly to become a binary, and drawing a companion mass from the distribution 
$0.6 (m_1/m_2)^{-0.4}$ (with $m_1 \leq m_2$) \citep{Kouwenhoven2007}. The semi-major axis is then drawn from a distribution 
that is uniform in $\log a$ in the range 0.005-50 AU, and eccentricity is chosen from the thermal distribution.
The main difference between the initial model used here and the types of models presented in MOR15
is that the model presented here is much more extended, with a half-mass radius of 7.56 pc, compared
to the typical half-mass radii of around 2 pc for the models from our main study. 

We have converted the initial conditions used by Wang et al. into a format that can be used in CMC, which 
means our model is an identical star-by-star representation of the $N$-body model. Of course, to retain
spherical symmetry we needed to convert the star coordinates $x,\, y,$ and $z$ to a radial coordinate $r$, 
and the velocity vectors $v_x,\, v_y$, and $v_z$ to their corresponding radial and transverse vectors, 
$v_r$ and $v_t$.  All other initial properties, including stellar masses, positions, and binary properties, are identical.

\section{Comparison to Direct $N$-body} \label{MC_NB_Comparison}


In what follows, we
make direct comparisons between the two simulations up to a time of 12 Gyr.
In Section \ref{struct_evol}, we explore the evolution and dynamics of the BHs in both simulations, while in Section \ref{disagreement} we highlight the agreement (and disagreement) between the structural properties of both cluster models.

\begin{figure}
\centering
\includegraphics[trim=2cm 0cm 2cm 0cm,scale=0.6]{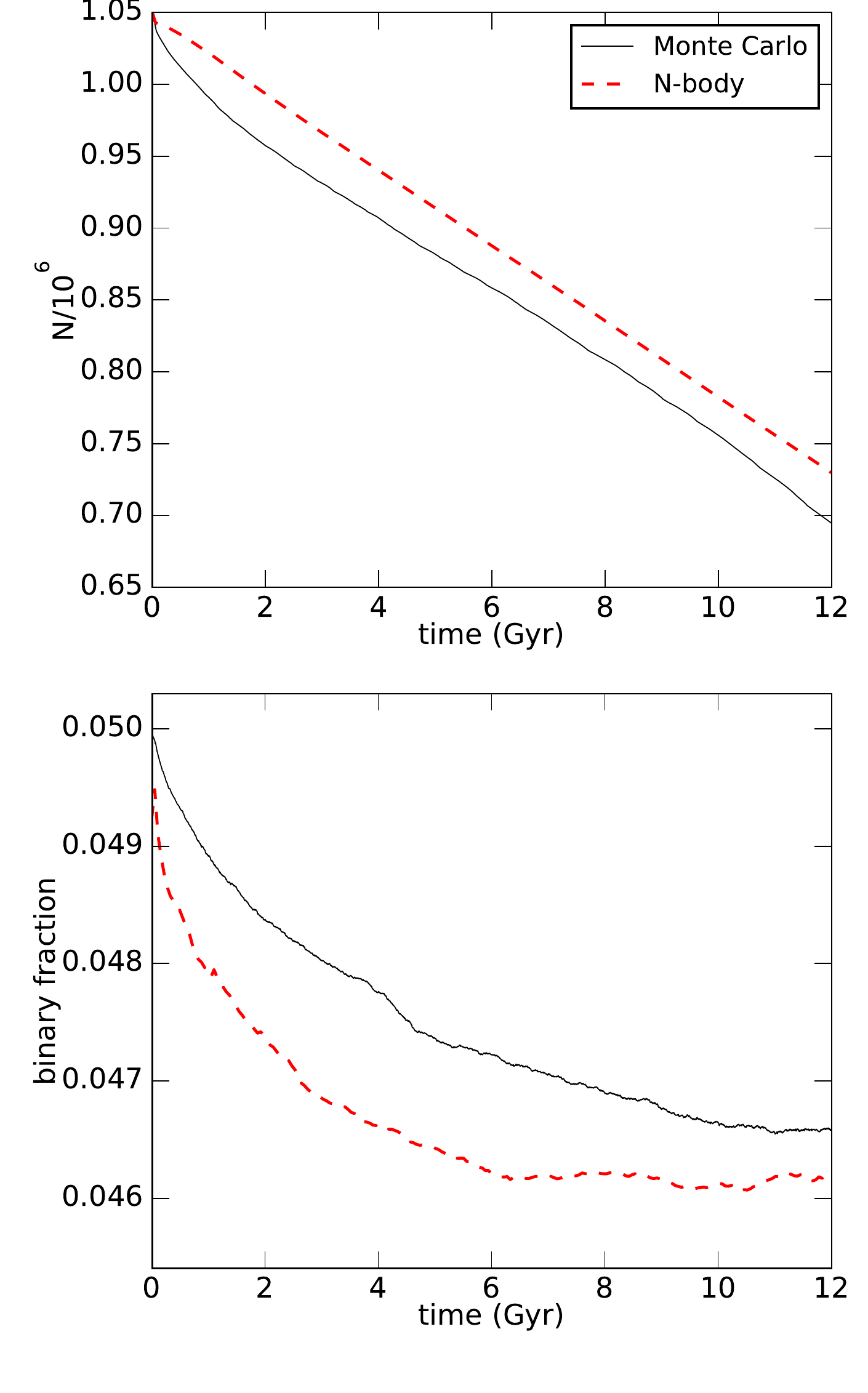}
 	 \caption{Top: Number of bound stars remaining in the cluster as a function of time for model \texttt{MC} 
	 (solid black) and model \texttt{NB} (dashed red). Each star is counted
	 individually, regardless of whether it is a part of a binary (or an even higher-order system, which is possible
	 in the direct $N$-body simulation). Bottom: binary fraction
	 as a function of time, with the same color scheme as above.}
	\label{fig:N_fb}
\end{figure}

\subsection{Black Hole Retention and Ejection} \label{struct_evol}

We start with our most exciting result: the direct $N$-body simulation confirms the findings of MOR15
that large numbers of BHs can be retained for many Gyr in old GCs. 	
At the end of the simulation, there are 1085 BHs remaining in model \texttt{MC} and 1036 in model \texttt{NB}. 
Figure \ref{fig:bhmass_hist_final} shows the initial and final distributions of BHs masses for the two models, which are in nearly perfect agreement.  Dynamically, the two models produce extremely similar results, with model \texttt{MC} ejecting 336 BHs and model \texttt{NB} ejecting 400 BHs from the population that is retained initially, and, in agreement with MOR15, ejecting the most massive BHs first.  We do note that model \texttt{NB}
contains a handful of BHs with masses above $35\, M_\odot$, whereas model \texttt{MC} has none. 
These abnormally massive BHs are formed as a result of a minor bug in the $N$-body treatment of merged stellar binaries, which was discovered after the simulation had begun.

On the bottom panels in the same figure we show the cumulative radial distributions of the BHs (solid curves) 
and the non-BHs (dotted curves) after BH formation and at the end of the simulation (solid curves). The two models show excellent agreement in the radial distributions
of both star types, with only slight disagreement growing in the outskirts of the cluster models for the non-BHs.
The crude tidal treatment in the MC code is not expected to reproduce the more accurate three-dimensional 
treatment in the direct $N$-body code. It is clear that the BHs, while more centrally concentrated than the non-BHs, 
are still quite spread out over the cluster. About half of the BHs lie outside of the inner 4 pc, and about 10\% lie 
beyond 10 pc.

In Figure \ref{fig:Nbh_comp_LW} we show the total number of BHs retained in each model as a function of time. 
The agreement is excellent throughout the formation of nearly all of the BHs, with the \texttt{MC} model forming and retaining an initial population of 1421 BHs and the \texttt{NB} model forming 1436 BHs.  
By about 20 Myr, after all BHs have formed and the ones with large natal kicks have been ejected
from the cluster, the number of BHs in each model levels out to the initially retained numbers given above.
Over time the number of BHs decreases slowly in each model at a similar rate, with the \texttt{NB} model ejecting 64 more BHs than the \texttt{MC} model by 12 Gyr.  Part of this increase can be attributed to the more-massive BHs (35-55 $M_{\odot}$) found in the $N$-body simulation which will be ejected faster than the $\sim 25 M_{\odot}$ BHs.  However, the 18 erroneously large BHs cannot account for the full discrepancy in the number of ejected BHs between the two models.  

Finally, we compare the properties of the ejected BH binaries.  In Figure
\ref{fig:bhformation}, we show the number of BH binaries ejected by each model
over time.  As the \texttt{NB} model is ejecting BHs more quickly than the
\texttt{MC} model, the rate of binary BH ejection is also larger in the $N$-body
model (54 versus 41 binaries).  Here, about half of these ejected binaries can
be attributed to binaries whose components are abnormally large.  Even then,
this suggests that the Monte Carlo method may underestimate the production rate
of binary BHs.  We also compare the component masses, eccentricities, and
semi-major axes of the ejected binaries (the middle and bottom panels of the same figure).  In this instance, we find very good agreement between the \texttt{NB} and \texttt{MC} models.  Such good agreement is to be expected, as both codes model binary hardening and partner exchanges (via three and four-body encounters) using a direct $N$-body integration.

\subsection{Structural Cluster Properties} \label{disagreement}

Next we compare the overall structural evolution of the two models.
In Figure \ref{fig:N_fb} we show the number of bound stars remaining in the cluster (top) and the binary fraction (bottom) as a function of time, with model \texttt{MC} shown in black (solid line) and model \texttt{NB} in red (dashed line). Here the number of bound stars is determined by counting every star individually (i.e., a binary counts as two stars). By the final time of 12~Gyr, both models have lost roughly
30\% of the initial $1.05 \times 10^6$ stars. The two models agree to within 5\% (with models \texttt{MC} and \texttt{NB} having $6.95 \times 10^5$ and $7.29\times10^5$ stars, respectively). The $N$-body code uses a radial tidal criterion 
such that any star that goes beyond twice the cluster tidal radius is removed. In contrast, the MC code relies on an energy criterion, which has shown to produce better agreement with direct $N$-body than a radial criterion, but has still been
observed to strip stars on a slightly shorter timescale \citep{Chatterjee2010}. The binary fraction in the models shows even better agreement to within about 1\%, decreasing rather slowly with time from $f_b$=0.05 initially down to $f_b$=0.0465 in \texttt{MC} and $f_b$=0.0462 in model \texttt{NB}.

\begin{figure*}[
\centering
\includegraphics[trim=0cm 0cm 0cm 0cm,scale=0.8]{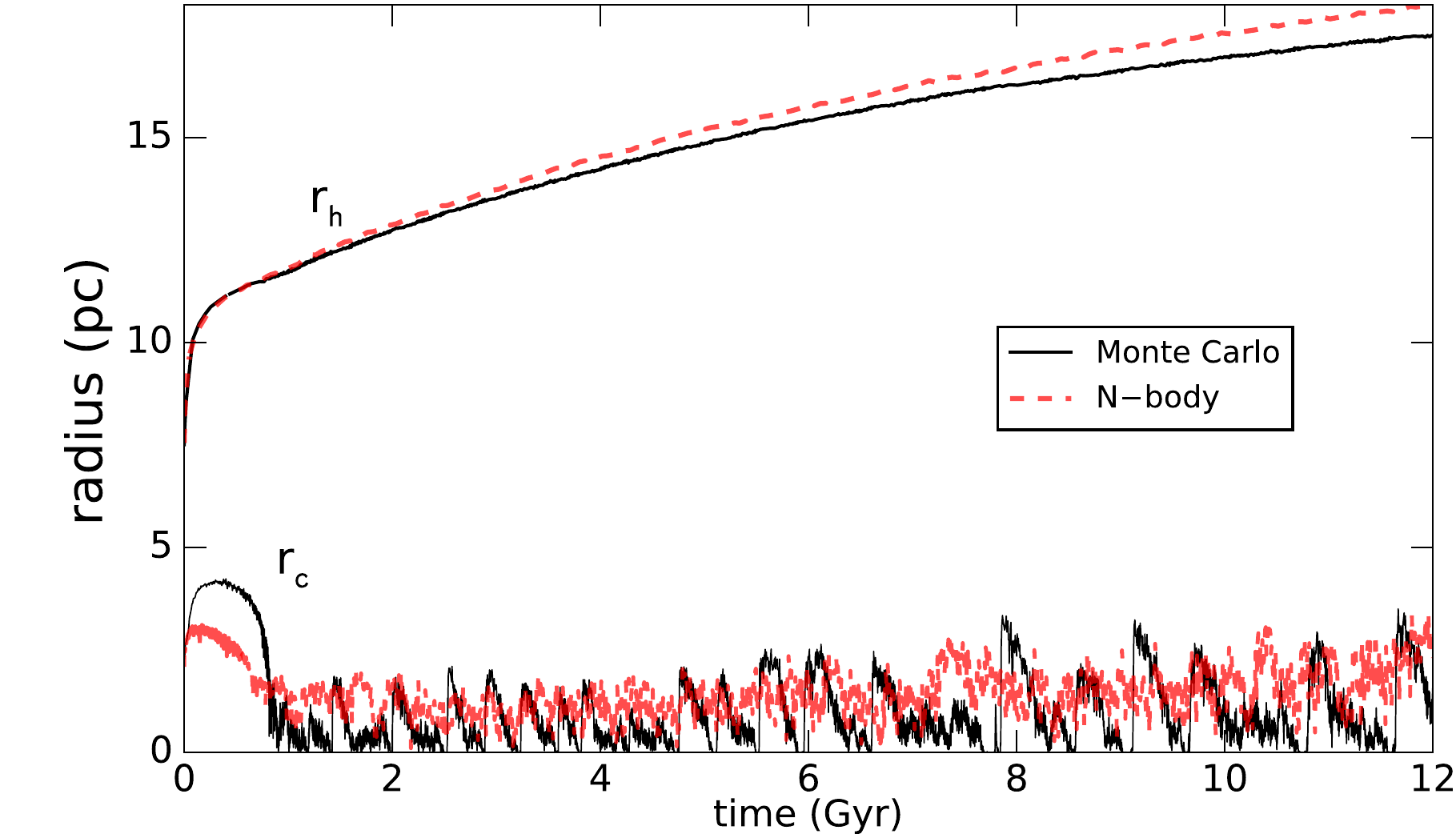}

 	 \caption{Comparison of the time evolution of the core radius ($r_{\rm c}$) and 
			half-mass radius ($r_{\rm h}$) between \texttt{MC} (solid black line) 
			and \texttt{NB} (dashed red line) up to the current time of the $N$-body 
			simulation. The two models show good agreement in the evolution of
			 the half-mass radius (to within $\sim 1$ pc), and fairly good agreement in the core radius evolution. 
			 We note that the core-collapses in model \texttt{MC} go deeper
			  than those seen in \texttt{NB} (see Figure \ref{fig:LagradLog} for a more detailed
			  look at what is happening in the central region).
			  }
	\label{fig:RcRh_linear}
\end{figure*}

In Figure \ref{fig:RcRh_linear} we show the time evolution of the core radius, $r_c$, and the half-mass 
radius, $r_h$, again with model \texttt{MC} in black and model \texttt{NB} in red.
 Here we use the standard $N$-body definition of the core radius from \cite{Casertano1985}, 
 which is a density-weighted estimate of the average of the star positions.
This theoretical core radius is of course unrelated to that which an observer would
calculate, since it depends on mass rather than luminosity. The half-mass radius is
the radius that encloses half of the total cluster mass.
We find nearly perfect agreement between the two models in the evolution of $r_h$ for the entire 
span of the simulations.
The agreement in the core radius is also very good, although at early times (within a few hundred Myr)
the core radius in model \texttt{MC} expands slightly more than that of model \texttt{NB}. 
This is likely caused by subtle differences in our stellar evolution routines (e.g. wind mass loss), which have both
been modified from the original publicly-available SSE and BSE codes (\citealt{Hurley2000, Hurley2002}).
These differences can be difficult to track down, and therefore we leave this topic for future investigations. 
In any case, since the core radii in both models have contracted down to a very similar value within a Gyr, 
and then remain in good agreement up to the end, this slight disagreement early on
appears to have very little influence on long-term evolution of the core radius. 

\begin{figure*}
\centering
\includegraphics[trim=0cm 0cm 0cm 0cm,scale=0.8]{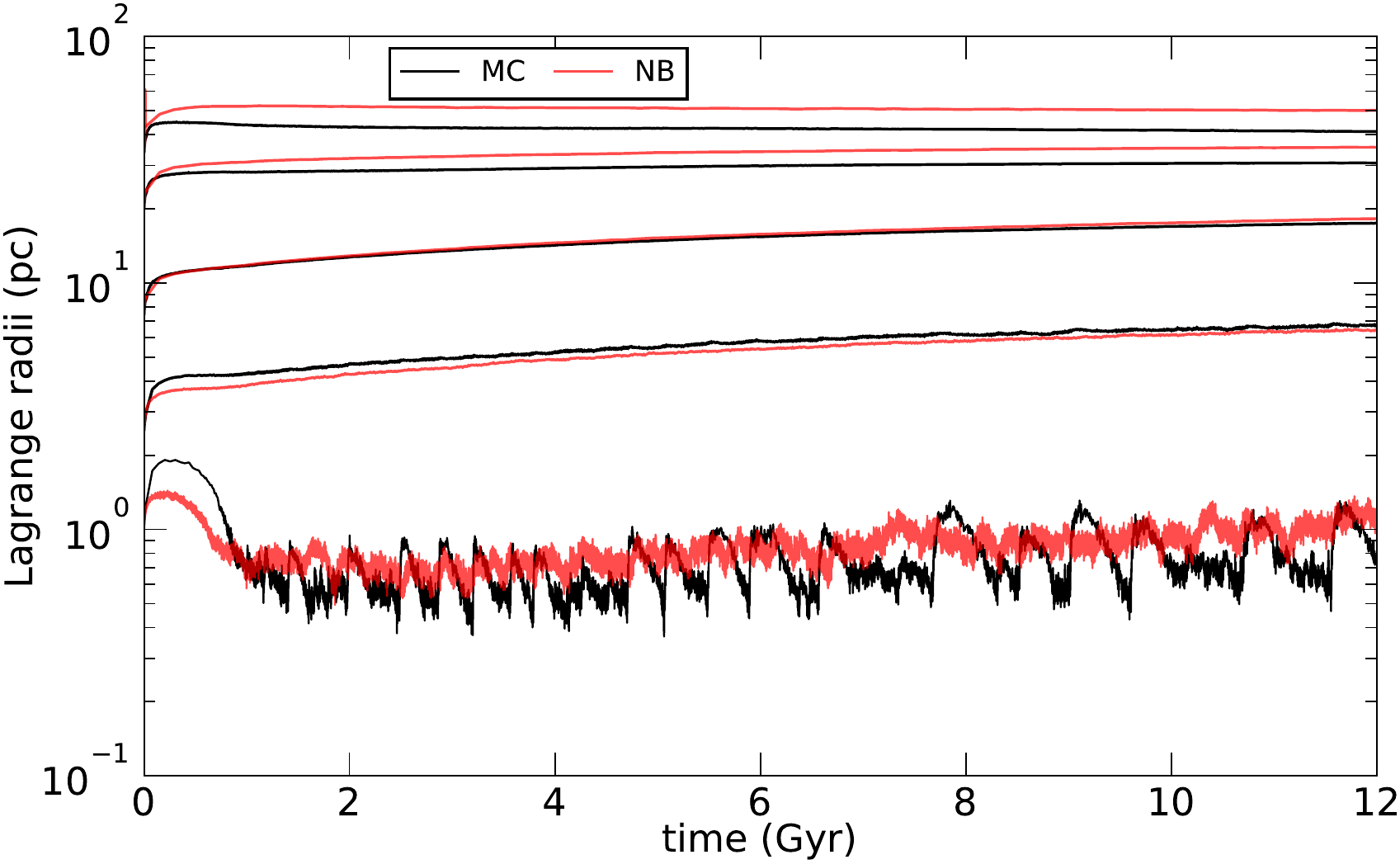}
 	 \caption{Comparison of the Lagrange radii for models \texttt{MC} (black) 
	 	and \texttt{NB} (red). The pairs of black and red curves show the radii enclosing 
		1\%, 10\%, 50\%, 90\% and 99\% (bottom to top) of the total cluster mass for 
		the two respective models over the entire simulation.
		}
	\label{fig:LagradAllstars}
\end{figure*}

In Figure \ref{fig:LagradAllstars} we show a comparison of the Lagrange radii for the two models, with 
model \texttt{MC} in black and model \texttt{NB} in red. From bottom to top, the pairs of black and red 
lines correspond to the radii enclosing 1\%, 10\%, 50\%, 90\% and 99\% of the total mass in the cluster.
Overall we find strong agreement between the two models. In the outer parts of the cluster (beyond
a few tens of pc), model \texttt{NB} expands slightly more than in model \texttt{MC} early on, 
and the memory of this seems to last through the end of the simulations. Again, we do not necessarily
expect the two codes to produce identical results because of the very different tidal stripping prescriptions employed.
More importantly, the 10\% and 50\% Lagrange radii agree nearly perfectly. The 1\% radius in model \texttt{MC} is noisier 
 and displays slightly deeper collapses than that of model \texttt{NB}, although the outer envelope of the 
two curves matches very well. The behavior of the 1\% radius is similar to that of the core radius shown in 
Figure \ref{fig:RcRh_linear}. Since the MC approach is not designed to handle the physics of small 
numbers of objects, it is actually quite impressive that the behavior on these small scales agrees as 
well as it does with the direct $N$-body simulation.

%
%

\begin{figure*}
\centering
\includegraphics[trim=0cm 0cm 0cm 0cm,scale=0.8]{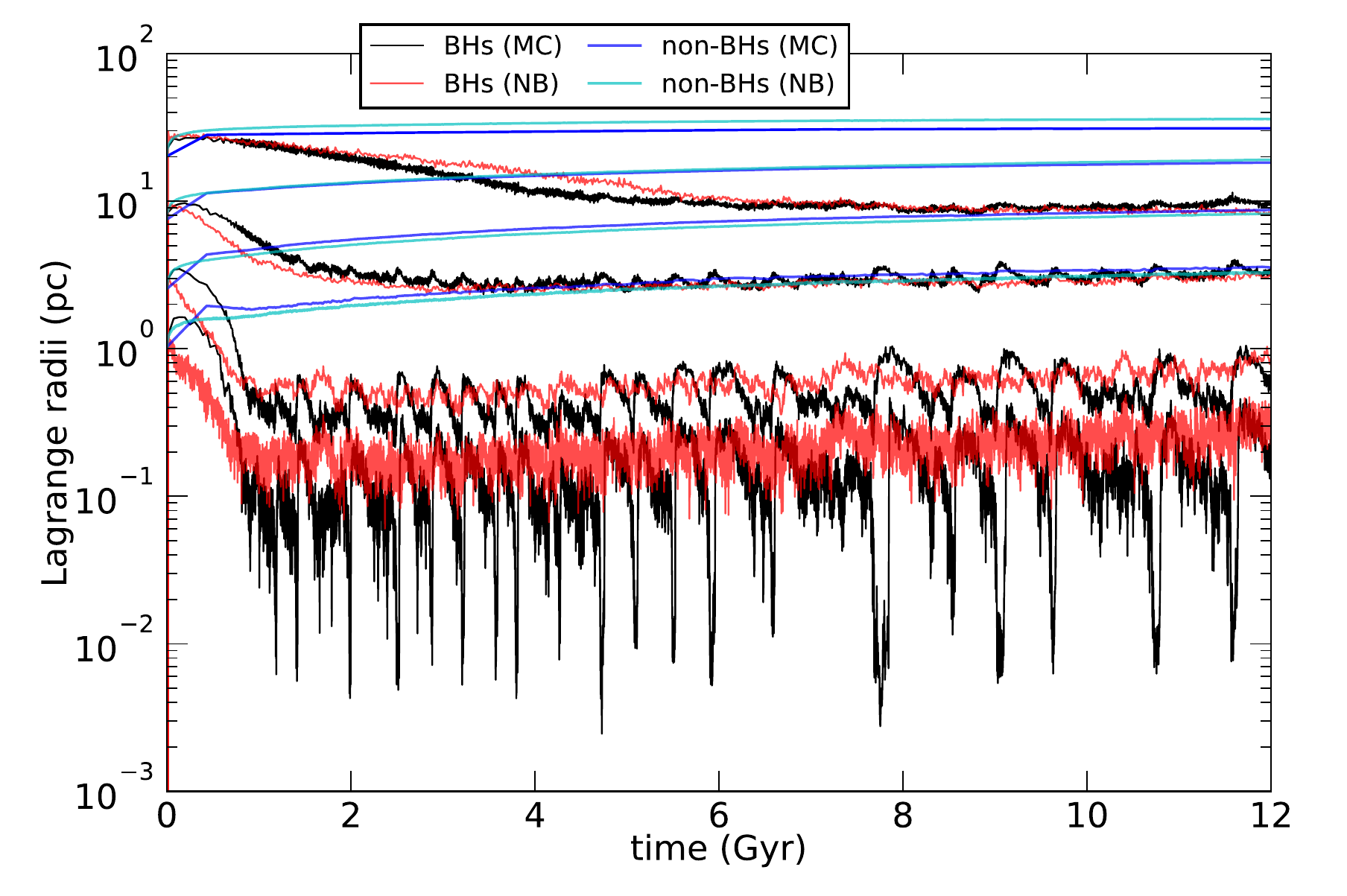}
 	 \caption{ Lagrange radii for the BHs and the non-BHs separately. 
	 	The black (\texttt{MC}) and red (\texttt{NB}) curves show the radii enclosing 1\%, 10\%, 50\% and 90\%
		of the \emph{BH} mass.
		Similarly, the blue (\texttt{MC}) and cyan (\texttt{NB}) curves show the radii enclosing 
		the same fractions of the \emph{non-BH} mass,
		Here we can see that it is primarily the BHs that are driving the core oscillations
		shown in Figure \ref{fig:RcRh_linear}.
		The 10\%, 50\% and 90\% BH Lagrange radii for both models agree extremely well, although
		in model \texttt{MC} the 10\% radius is a bit noisier. The 1\% BH Lagrange radius in model
		\texttt{MC} collapses significantly deeper than in model \texttt{NB}, however this bin typically
		contains only about 10 BHs.
		The BHs remain very well mixed with other stars (even at 12 Gyr, 90\% of the BH mass 
		is spread throughout the $\sim10^5$ stars comprising the inner 10\% of non-BH mass).
		}
	\label{fig:LagradLog}
\end{figure*}


In Figure \ref{fig:LagradLog} we again show the Lagrange radii, only this time we separate the BHs from the 
other types of objects in order to better understand the behavior of the BHs. With model \texttt{MC} in black
and model \texttt{NB} in red, we show (from bottom to top) the 1\%, 10\%, 50\% and 90\% radii for the BHs,
and the same Lagrange radii for the non-BHs in blue for model \texttt{MC} and cyan for model \texttt{NB}.
We can see that the innermost 1\% of the BH mass in both models 
participates in oscillations similar to those seen in the core radii in Figure \ref{fig:RcRh_linear} 
(i.e., deeper oscillations in the MC model, more shallow ones in the $N$-body), but these oscillations are not
seen in the non-BH population. 
This tells us that these collapses are primarily driven by the BHs.
This innermost 1\% mass bin for the BHs consists of only $\sim10$ objects, so it is unreasonable to expect
the MC code to treat this subsystem perfectly.
In all but the innermost 1\% BH mass, we find excellent agreement between the two models in the 
radial distributions of both species (BHs and non-BHs) as a function of time. 
The outermost 90\% Lagrange radii for the BHs falls at about 10 pc for both models at the end of the 
simulations, and the 90\% radius for the non-BHs lies just a bit further out at around 30 pc. 
Thus the BHs remain quite spread out 
in the cluster, well-mixed with the other objects, all the way to the end of the simulation.
Despite the deeper collapses of the central BH subsystem in model \text{MC}, 
the fact that the BH ejection rate agrees so well provides encouraging evidence that BH dynamics
and retention is not necessarily governed by the depth of the BH-driven core oscillations. 
The only noticeable change in the BH distributions between 100 Myr and 12 Gyr (Figure \ref{fig:bhmass_hist_final}) is at the high
mass end, with the heavy BHs being among the first to be ejected. The two models agree very well in this regard.
Thus, not only do the \emph{numbers} of ejected BHs agree, but so too do the
masses of the ejected BHs.


\section{Discussion} \label{Discussion}

We have found that our MC approach and the new code NBODY6++GPU produce similar evolution 
for a million-star cluster over 12~Gyr, both in terms of global structural properties and the dynamical
evolution of the BH populations. Despite the significant agreement in the dynamics,  we are still interested in exploring 
why the BHs in model \texttt{MC} experience 
deeper collapses than in model \texttt{NB}, as this may help us better understand the nature 
of the interactions between BHs and the rest of the cluster. 
One potential cause for the deeper collapses in the MC model would be the presence of more massive 
BHs than in the $N$-body model. MOR15 found that it is always the \emph{most massive} BHs
that are found near the center and driving the deep core oscillations. As a result, they are also the first to be
ejected. However, we actually find the \emph{opposite} to be true: for the first Gyr, the $N$-body model contains 
about ten BHs in the mass range $35-55\, M_\odot$, whereas the MC model contains none in this range.
By 4 Gyr, these heavy BHs have been ejected from model \texttt{NB}. 
In any case, it is clear that the deep collapses in the MC model cannot be explained by the presence of more massive BHs. Furthermore, we saw in Figure \ref{fig:bhmass_hist_final} that, besides the handful of BHs with basses between $35-55\, M_\odot$, the overall distribution of BH masses retained in the two models agrees fairly well throughout the entire simulation.

Another possible explanation for the difference in the collapse depths could be
differences in the population of BH \emph{binaries} that arise from the simplified three-body binary 
formation prescription used in CMC.
In our standard prescription, we allow 
\emph{only BHs} to form binaries through three-body encounters, a choice motivated by previous studies suggesting
that for non-compact stars, the high densities required to form three-body binaries would instead lead to physical
collisions of these stars, rather than forming binaries (\citealt{Chernoff1996}), and is thus never important.  
This means that in model \texttt{MC} we are suppressing the creation of BH-non-BH binaries by three-body interactions, and
our population of BH-non-BH binaries is limited to the number of BHs that can exchange into existing 
binaries through binary interactions. The smaller number of BH binaries (of any type) in the MC model may be 
responsible for the deep core collapses which would be overturned much earlier in the collapse if we had a 
larger number of BH binaries. Alternatively, the collapse depths could be sensitive to the \emph{hardness} of the three-body binaries that are formed, or the actual \emph{rate} of binary formation, both of which are determined in part by the
parameter $\eta_{\rm min}$, the minimum hardness of the binaries that are allowed to form in the simple 
prescription (see MOR15 for the full details of the procedure).
This parameter affects binary formation in two different ways: first, 
$\eta_{\rm min}$ enters the rate equation, so the rate of binary formation in a given timestep with this
minimum hardness $\Gamma (\eta \geq \eta_{\rm min}) \sim \eta_{\rm min}^{-3.5}$, meaning that it is easier 
to form softer binaries (smaller $\eta$). Secondly, if our procedure determines that a binary should form, 
then we choose the \emph{actual} hardness for the binary from a distribution according to the differential
rate, d$\Gamma$/d$\eta$, with lower limit $\eta_{\rm min}$. This function falls off rapidly for increasing values
of $\eta$, so a lower limit will tend to result in significantly softer binaries being formed. In model \texttt{MC}, as 
well as in the models presented in MOR15, we have strictly used a hardness cutoff $\eta_{\rm min}=5$.
For full details of the three-body binary formation prescription employed in CMC, see MOR15.

To test the effect of modifications to our binary formation procedure we performed two additional
simulations identical to \texttt{MC}, except for a slight modification to our three-body binary treatment.
We have performed one simulation in which all stars, including non-BHs, are allowed to participate in three-body
binary formation. We find, however, that this change has essentially no effect on the BH-driven collapses nor on the 
BH evaporation rate. Next we reduced $\eta_{\rm min}$ to 0.5 (a factor of ten smaller than used in model \texttt{MC}).
In this case we saw a slight reduction in the collapse depths, but the number of retained BHs remained unaffected.
It seems that when softer binaries are allowed to form, which also increases the overall rate of production,
the BH-driven collapses can be overturned slightly sooner, and the very deepest collapses are avoided. 
While these modifications to the simple treatment of binary formation do not seem to have much of an effect
on overall cluster evolution or on the retention of BHs, it may very well affect the details of the BH-binary
populations, such as the production of BH-X-ray binaries and tight BH-BH binaries.
It would be possible to study these binary populations in greater detail, and perhaps even use the direct $N$-body
model to calibrate the three-body binary procedure used in CMC, but we leave this for a future study.


In contrast to the MC models presented in MOR15, the initial model used in this study is much 
more extended. Other than slight differences in the initial binary populations, the rest of the initial model details
and the physics implementation remains
the same. The behavior of the core radius and the inner BH Lagrange 
radii in model \texttt{MC} is nonetheless very similar to that seen in the much more compact MC models presented in MOR15. Also, the order in which BHs are ejected, from heavy to light, agrees with the findings
of MOR15. The main difference in the evolution of model \texttt{MC} presented here is that, being much more extended, 
the cluster has a significantly longer relaxation timescale, and therefore it processes (and thus ejects) BHs much more 
slowly than more compact models with similar $N$ from MOR15, which have typically ejected $\sim$ 1000s of BHs on the same physical timescale.

\section{Summary and Conclusions} \label{SummaryConclusions}

In this study we have compared the results of realistic large-$N$ star cluster simulations
created using two very different techniques: an orbit-averaged MC approach
(CMC) and new direct $N$-body code (NBODY6++GPU). 
In terms of overall dynamical structure and evolution, as well as the dynamics of the BHs and long-term 
BH retention, we find quite remarkable agreement between the two techniques.
Most notably, the direct $N$-body model confirms the finding of MOR15 that very large numbers of BHs
($\sim1000$) can be retained for $\sim10$~Gyr timescales in old GCs.

We see slight differences in the evolution of the
innermost 1\% BH Lagrange radius: model \texttt{MC} displays core oscillations driven by the 
heaviest BHs, which do not occur in model \texttt{NB}. However, this disagreement does not seem 
to have any significant impact
on the overall evolution of the cluster models. For example, the half-mass radius evolution of model 
\texttt{MC} models agrees with that of the $N$-body model. Additionally, both models eject BHs at 
roughly the same rate, losing just over 300 of the initially retained BHs over the entire simulation and thus
still retaining more than a thousand BHs at the end of the simulation. 
This verifies that, while our approximate MC approach may not be capable of treating the physics of the small number of BHs in the core perfectly, our technique is nonetheless capable of reproducing the bulk evolution of the cluster as seen in the 
$N$-body model rather well, making it a viable technique for modeling the dynamical evolution of realistic clusters
containing large numbers of stellar-mass BHs. With our parallel CMC code, a cluster model like the one studied here
can be completed (evolved up to $12$~Gyr) within about a day. The same model computed with NBODY6++GPU requires more than 6 months to complete.

MOR15 argued that the long-term survivability of BHs in clusters
relies on the fact that the BHs mostly \emph{avoid} the Spitzer instability, in contrast to what has often been assumed.
The lack of BH-driven collapses in the $N$-body model presented here provides evidence that the BHs are 
even \emph{less} susceptible to the Spitzer instability than predicted previously by MOR15.
This would provide strong
 support for the main conclusion from MOR15, namely that \emph{if} large numbers of BHs are retained initially 
 (as they are under standard assumptions regarding star cluster initial conditions and BH formation processes), then 
 many will be retained at 12 Gyr, especially for a model with a large initial virial radius (as studied in this work).

\section{Acknowledgments}

This work was supported by NSF Grant AST-1312945 and NASA Grant NNX14AP92G. All
computations using CMC were performed on Northwestern University's HPC cluster
Quest.  CR was supported by an NSF GRFP Fellowship, award DGE-0824162.  MM was
supported by an NSF GK-12 Fellowship, award DGE-0948017. W. L.
acknowledge the support by the Silk Road Project at the National Astronomical
Observatories of China (NAOC, http://silkroad.bao.ac.cn), the
Max-Planck-Institute for Astrophysics and the Max Planck Computing and Data
Facility (MPCDF, http://www.mpcdf.mpg.de/) in Garching, Germany. All direct $N$-body simulations were run on their Hydra GPU cluster.


\bibliographystyle{mnras} 
\bibliography{bibtex}


\bsp	
\label{lastpage}
\end{document}